\documentclass[twocolumn]{aastex631}

\usepackage{amsmath}
\usepackage{amssymb}

\usepackage[T1]{fontenc}

\newcommand{\ketju}{{KETJU}}

\newcommand{\gadget}{GADGET-3}

\newcommand{\dd}[1]{\mathop{}\!\mathrm{d}#1}
\newcommand{\dv}[2]{\frac{\dd#1}{\dd{#2}}} 

\newcommand{\pn}[1]{#1PN}

\renewcommand{\edit}[2]{#2}

\begin{document}
\title{
Resolving the Complex Evolution of a Supermassive Black Hole Triplet in a Cosmological Simulation  
}

\shorttitle{
SMBH Triplet Evolution in a Zoom Simulation
}

\author[0000-0001-5721-9335]{Matias Mannerkoski}
\affiliation{
Department of Physics,
Gustaf H\"allstr\"omin katu 2, FI-00014, University of Helsinki, Finland
}

\author[0000-0001-8741-8263]{Peter H. Johansson}
\affiliation{
Department of Physics,
Gustaf H\"allstr\"omin katu 2, FI-00014, University of Helsinki, Finland
}

\author[0000-0001-8789-2571]{Antti Rantala}
\affiliation{
Max-Planck-Institut f\"ur Astrophysik, Karl-Schwarzchild-Str 1,
D-85748 Garching, Germany
}

\author[0000-0002-7314-2558]{Thorsten Naab}
\affiliation{
Max-Planck-Institut f\"ur Astrophysik, Karl-Schwarzchild-Str 1,
D-85748 Garching, Germany
}

\author[0000-0001-7075-6098]{Shihong Liao}
\affiliation{
Department of Physics,
Gustaf H\"allstr\"omin katu 2, FI-00014, University of Helsinki, Finland
}

\correspondingauthor{Matias Mannerkoski}
\email{matias.mannerkoski@helsinki.fi}

\begin{abstract}

We present here a self-consistent cosmological zoom-in simulation of a triple supermassive black hole (SMBH) system forming in a complex
multiple galaxy merger. The simulation is run with an updated version of our code \ketju{}, which is able to follow the 
motion of SMBHs down to separations of tens of Schwarzschild radii while simultaneously modeling the large-scale astrophysical 
processes in the surrounding galaxies, such as gas cooling, star formation, and stellar and AGN feedback. Our simulation
produces initially a SMBH binary system for which the hardening process is interrupted by the late arrival of a third SMBH. 
The \ketju{} code is able to accurately model the complex behavior occurring in such a triple SMBH system, including the ejection of one SMBH to a kiloparsec-scale orbit in the galaxy
due to strong three-body interactions as well as Lidov--Kozai oscillations suppressed by relativistic precession when the SMBHs are in a hierarchical configuration.
One pair of SMBHs merges $\sim 3\,\mathrm{Gyr}$ after the initial galaxy merger, while the remaining binary is at a parsec-scale separation when the simulation ends at redshift $z=0$.
We also show that \ketju{} can capture the effects of the SMBH binaries and triplets on the surrounding stellar population, which can affect the binary
merger timescales as the stellar density in the system evolves. Our results demonstrate the importance of dynamically resolving the complex behavior 
of multiple SMBHs in galactic mergers, as such systems cannot be readily modeled using \edit1{simple orbit-averaged} semi-analytic models. 

\end{abstract}

\section{Introduction}

Supermassive black holes (SMBHs) with masses in the range of $M_{\rm BH}=10^{6}\textnormal{--}10^{10} M_{\odot}$
are found in the centers of all massive galaxies \citep{kormendy2013}.
In the $\Lambda$CDM hierarchical model galaxies grow through mergers and gas accretion, with the
coalescence of SMBHs in galactic mergers proceeding through three stages \citep{begelman1980}.
First, the separation between the SMBHs shrinks from the initial kiloparsec scale due to dynamical
friction \citep{1943ApJ....97..255C} from the surrounding stars and gas in the galaxy until the
SMBHs form a bound binary with a typical separation of $a\sim {1\textnormal{--}10 \,\mathrm{pc}}$.
From there the binary will further shrink (`harden') due to scattering of individual stars that carry away energy and
angular momentum \citep{1980AJ.....85.1281H}.
Finally, at sub-parsec scales gravitational wave (GW) emission becomes the dominant
mechanism for energy loss and drives the SMBH binary to coalescence \citep{1963PhRv..131..435P, 1964PhRv..136.1224P}. 

Given a suitable galactic environment where the lifetime of the SMBH binary exceeds the time between galactic mergers,
systems that include multiple interacting SMBHs may form (e.g. \citealt{2007MNRAS.377..957H}).
The triplet is the simplest multiple SMBH configuration,
and the dynamics of such systems have been extensively studied in 
isolated simulations\edit1{, including a semi-analytic treatment of the stellar environment
\citep{2007MNRAS.377..957H,2016MNRAS.461.4419B,2019MNRAS.486.4044B}.}
In addition, there is now increasingly strong evidence that 
such systems are relatively commonplace, as several triplet SMBHs have been observed in the 
local Universe \citep{2014Natur.511...57D,2019ApJ...883..167P,2019ApJ...887...90L,2020A&A...633A..79K}. 

Modeling the entire SMBH coalescence process beyond the formation of a bound binary has not previously been possible in a full cosmological simulation 
due to the inability of simultaneously modeling the small-scale dynamics and global galactic-scale processes
in simulations that include gravitational force softening  \citep{2017MNRAS.464.3131K,2018MNRAS.473.3410R}. 
Some improvements on the SMBH behavior at kiloparsec scales have been achieved with the addition of subgrid
models of the unresolved dynamical friction contribution \edit1{\citep{2015MNRAS.451.1868T, 2019MNRAS.486..101P}.}
However, the parsec-scale dynamics has in general been modeled by post-processing the
simulations using semi-analytic methods \edit1{based on orbit-averaged equations} \citep{2017MNRAS.464.3131K,2017MNRAS.471.4508K}
or by resimulating the core regions of the merged galaxies using an
altogether separate N-body code \citep{2016ApJ...828...73K}.
Both of these approaches break the coupling of the small-scale SMBH dynamics
with the global simulation, which affects the ability to self-consistently model
the evolution of the stellar structure of the galaxy and 
may have important consequences for both the merger timescales of the SMBHs and
the structure of the final galaxy \citep{2018ApJ...864..113R}.

In this Letter we present a self-consistent cosmological zoom-in simulation of a triple SMBH system forming in a complex 
multiple galaxy merger at redshift $z\sim 0.5$.
The simulation is run with our \ketju{} code, which is capable of
following the motion of SMBHs down to separations of tens of Schwarzschild radii
while simultaneously modeling the large-scale processes in the surrounding galaxies.

\section{Numerical Simulations}
\subsection{The KETJU Code}

The simulations are run using the \ketju{} code \citep{2017ApJ...840...53R}, which is
an extension of the widely used \gadget{} code \citep{2005MNRAS.364.1105S}.
In the \ketju{} code
the dynamics of SMBHs and the stars in a small region around them are integrated with an algorithmically
regularized integrator, whereas the dynamics of the remaining particles is computed with the GADGET-3 leapfrog
using the tree-PM force calculation method. The application of an algorithmically
regularized integrator enables the accurate modeling of dynamical friction on SMBHs and SMBH binary
hardening,
provided that the SMBH mass to stellar particle mass ratio is large enough.
A mass ratio of $\sim 500 \textnormal{--} 1000$ has been observed to give converged
results with only a weak resolution dependence \citep{2017ApJ...840...53R}.

In this paper we have replaced the regularized AR-CHAIN \citep{2008AJ....135.2398M} integrator used in
the first \ketju{} version \citep{2017ApJ...840...53R} with the new MSTAR integrator \citep{2020MNRAS.492.4131R},
which has a significantly improved parallelization scheme and an improved interface with the main \gadget{} code. 
Together these improvements allow for
simulations containing up to $\sim 10^4$ particles in the regularized regions without the computational cost becoming prohibitive, which
is a significant improvement on the previous \ketju{} studies \citep{2017ApJ...840...53R,2018ApJ...864..113R,2019ApJ...872L..17R,2019ApJ...887...35M}.

The integration within each regularized region is performed in physical center-of-mass coordinates, converted from (to) the comoving coordinates used in the main
integrator at the start (end) of each integration, while the center of mass of
the system is propagated in comoving coordinates.
This correctly captures the motion of the system in an expanding universe.
We set the relative per step error tolerance of the integrator to $\eta=10^{-8}$
in order to ensure accurate evolution also in the GW dominated regime.
To model the effects of general relativity on the motion of
the SMBHs, \ketju{} includes post-Newtonian (PN) correction terms up to order 3.5 between
each pair of SMBHs \citep{2004PhRvD..69j4021M}. 
However, mergers of SMBHs are currently implemented in \ketju{} only in a simplified fashion 
conserving the Newtonian linear and angular momentum as well as the total mass of the system,
with the SMBHs being merged at a separation of six Schwarzschild radii.

Finally, in order to avoid possible energy errors caused by interactions between stellar particles
just within and outside the rather large regularized regions, we now also employ gravitational
softening for the stellar particle interactions inside the regularized regions.
The introduction of stellar softening does not negatively
affect the accuracy of the SMBH dynamics as all interactions involving SMBHs are still non-softened.

\subsection{Hydrodynamics and Feedback}

Contrary to our earlier \ketju{} studies, the simulations presented here also include a gas component and both stellar and BH feedback.
The hydrodynamics are modeled using the modern SPHGal smoothed particle hydrodynamics (SPH) implementation \citep{2014MNRAS.443.1173H}, 
which employs a pressure-entropy formulation together with artificial conduction,
artificial viscosity and a Wendland $C^4$-kernel smoothed over 100 neighbors. 
\edit1{Currently, the small-scale gas dynamics around the SMBHs is not
resolved below the softening scale of the simulation.} 

For stellar physics and gas cooling we use metal-dependent cooling models
tracking 11 individual elements \citep{2005MNRAS.364..552S,2006MNRAS.371.1125S,2013MNRAS.434.3142A}.
Our star formation model stochastically converts gas particles
to stellar particles based on the local star formation timescale above a
critical hydrogen number density of $n_\mathrm{H} = {0.1\,\mathrm{cm}^{-3}}$.
Other features of the models include feedback on gas from supernovae and massive
stars and the production of metals through stellar chemical evolution \citep{2013MNRAS.434.3142A,2017MNRAS.468..751E}.

Galaxies with dark matter halo masses of  $M_{\rm DM}=10^{10} h^{-1} M_\odot$ are seeded with
SMBHs with masses of $M_{\bullet}=10^5 h^{-1} M_\odot$ \citep{2007MNRAS.380..877S}.
Black holes grow through accretion and merging, with the accretion 
modeled using a standard Bondi--Hoyle--Lyttleton prescription 
with an additional dimensionless multiplier $\alpha = 25$ to account for the limited spatial resolution \citep{2009ApJ...707L.184J}. 
The accretion rate is capped at the Eddington limit assuming a
radiative efficiency of $\epsilon_{r}=0.1$ and with 5\% of the radiated energy coupling to the surrounding gas as thermal energy \citep{2005MNRAS.361..776S}.
A drawback of this accretion model is that it does not properly model accretion onto SMBHs in a binary system.
However, this
shortcoming is not significant for the particular binary and triple SMBH systems that we are 
concentrating on in this study, as the gas surrounding the black holes during
the binary phase is very dilute and hence the corresponding accretion rates are very low.

Due to the mass ratio
requirement we only switch on the regularized dynamics after the SMBHs of interest have grown to sufficiently large masses.
Before the regularized dynamics are switched on, i.e. when using standard \gadget{}, the SMBHs are kept in the centers of their host galaxies using a simple
repositioning method \citep{2009ApJ...690..802J},
which allows them to grow to realistic masses due to merging and gas accretion.

\subsection{Initial Conditions and Simulations}

We perform a cosmological zoom-in simulation starting at a redshift of $z=50$ centered on
a massive dark matter halo with a virial mass of $M_{200} \sim {7.5 \times 10^{12} M_\odot}$ at $z=0$. 
The initial conditions for our simulation are generated with the 
MUSIC \citep{2011MNRAS.415.2101H} software package.
We use  the Planck 2018 cosmology \citep{2020A&A...641A...6P}: 
$\Omega_m = 0.315$, $\Omega_b = 0.0491$, $\Omega_\Lambda = 0.685$, 
$H_0 = {h \times 100\,\mathrm{km}\,\mathrm{s}^{-1}\,\mathrm{Mpc}^{-1}} ={67.4\,\mathrm{km}\,\mathrm{s}^{-1}\,\mathrm{Mpc}^{-1}}$,
$\sigma_8 = 0.81$ and $n_s = 0.965$. 

The target halo is selected from an initial run of a uniform dark matter only
box with a comoving side length of $100h^{-1}\,\mathrm{Mpc}$ and $256^3$ particles.
We then generate new initial conditions with 4 levels of refinement around the
Lagrangian volume of the target halo, so that the highest-resolution region
contains approximately a total of $\sim {2\times 200^3}$ particles, with an equal number
of gas and dark matter particles. This results in a dark
matter particle mass of $m_\mathrm{DM}={1.6 \times 10^6 M_\odot}$ and a gas particle
mass of $m_\mathrm{gas}={3 \times 10^5 M_\odot}$ in the high-resolution region. 

The gravitational softening lengths are initially fixed in comoving coordinates. 
Below redshift $z=9$ the softening lengths
are fixed in physical coordinates at values of $\epsilon_\mathrm{bar}={40 h^{-1} \,\mathrm{pc}}$
for stars and gas and $\epsilon_\mathrm{DM,high}={93 h^{-1}\,\mathrm{pc}}$ for high-resolution dark matter particles.
The low resolution boundary dark matter particles have correspondingly much
larger softening lengths of $\epsilon_\mathrm{DM,low}={5.96 h^{-1}\,\mathrm{kpc}}$.
The radii of the regularized regions were set to ${120 h^{-1}\,\mathrm{pc}}$, corresponding
to $3\times \epsilon_\mathrm{bar}$.

\section{Results}
\label{sec:results}

\subsection{Simulation Overview}

\begin{figure*}
\centering
\includegraphics{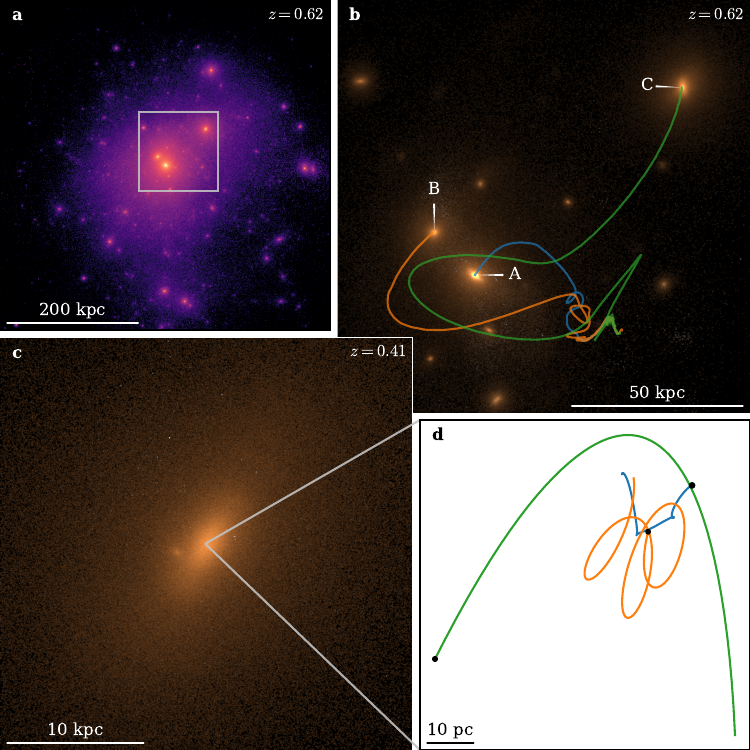}
\caption{
Overview of the simulation, showing the different physical scales modeled.
\textbf{a}: The simulated dark matter halo, showing the
projected mass density at redshift $z\approx 0.62$ corresponding to cosmic time
$t\approx 7.8\,\mathrm{Gyr}$ when the \ketju{} dynamics were switched on.
The box marks the region shown in panel b.
\textbf{b}: A BVR-image of the main galaxies A, B and C, with the colored lines showing the
subsequent trajectories of their central SMBHs until $z=0$. 
\textbf{c}: The galaxy formed after galaxies A, B and C have merged shown at
$z\approx 0.41$ ($t\approx 9.3\,\mathrm{Gyr}$).
\textbf{d}: The three SMBHs interacting in the center of the galaxy, showing
 sections of their trajectories spanning ${1\,\mathrm{Myr}}$.
}
\label{fig:overview}
\end{figure*}

We initially run the simulation with standard \gadget{} without the \ketju{}
SMBH dynamics enabled.
At redshift $z \approx 0.62$ the target halo hosts three massive galaxies
(A, B and C; \autoref{fig:overview}, panel b)
with stellar masses of 
$M_{*,A} = {1.4 \times 10^{11} M_\odot}$, $M_{*,B} = {5.4 \times 10^{10} M_\odot}$ 
and $M_{*,C} = {6.3 \times 10^{10} M_\odot}$  
(within ${30\,\mathrm{kpc}}$; all distances in this section are measured in physical coordinates). 
These galaxies host massive central SMBHs with masses of  $M_{\bullet,A} = {8.4\times 10^8 M_\odot}$,
$M_{\bullet,B} = {1.1 \times 10^8 M_\odot}$ and $ M_{\bullet,C} = {2.1 \times 10^8 M_\odot}$,
which are consistent with observed galaxies of similar masses \citep{kormendy2013}.
At this stage the mass ratio between these SMBHs and the
stellar particles (mean $m_\mathrm{part}\approx {2.5\times 10^{5} M_{\odot}}$)
is sufficiently large to allow for detailed dynamical modeling using \ketju{}. 
\edit1{
The corresponding gas fractions within $1\,\mathrm{kpc}$ from these SMBHs are
very low at $f_{\rm gas}=M_\mathrm{gas}/(M_\mathrm{gas}+M_*)\sim 10^{-4}$.}
From this point on, we continued the simulation using two different configurations,
with one simulation run using \ketju{} and the other run continued with standard
\gadget{} without SMBH repositioning to demonstrate the effects of our improved
SMBH dynamics compared to softened dynamics.
Both simulations were run until redshift $z=0$.

\subsection{Galaxy Mergers and SMBH Orbital Evolution}

\begin{figure*}
\centering
\includegraphics{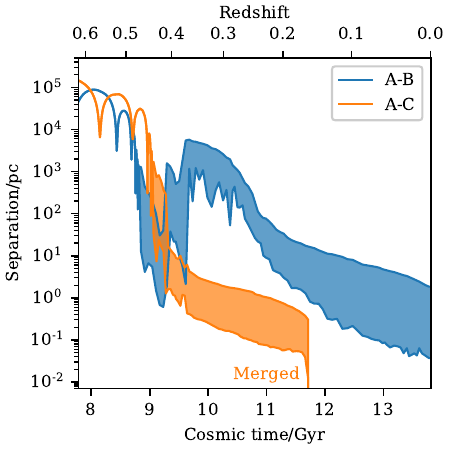}
\includegraphics{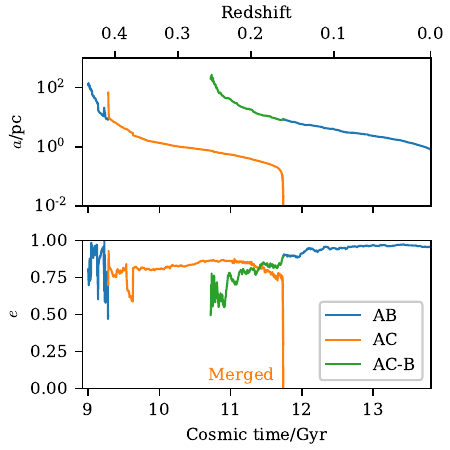}
\caption{
\edit1{
Left: The SMBH A-B and A-C separations over the \ketju{} simulation.
Shaded regions show the range of rapid oscillations.
}
Right: Evolution of the semimajor axis $a$ and eccentricity $e$ for the SMBHs in the
system.
Binaries are labeled by the letters corresponding to their constituent SMBHs
(e.g. AB is the binary consisting of SMBHs A and B), while AC-B denotes the
orbit of B around the AC binary in a hierarchical configuration.
The remnant of the AC binary merger is also labeled as A.
}
\label{fig:orbital_evolution}
\end{figure*}

\begin{figure}
\includegraphics{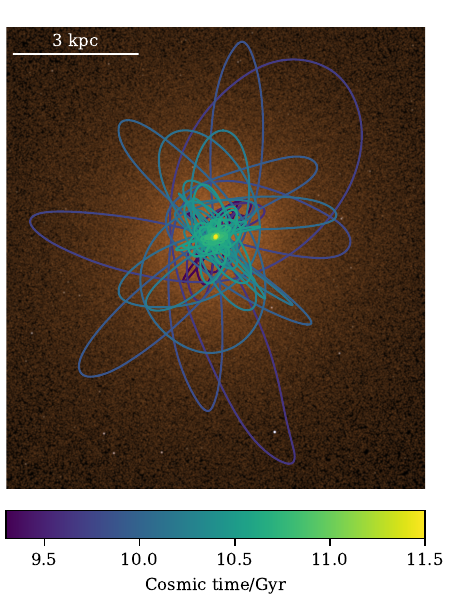}
\caption{
Orbit of the ejected SMBH-B between cosmic times 
$t={9.3\,\mathrm{Gyr}}$ $(z\sim 0.4)$ and $t={11.5\,\mathrm{Gyr}}$ $(z\sim 0.2)$
overlaid on the image of the galaxy.
}
\label{fig:ejected_bh_orbit}
\end{figure}

\begin{figure}
\centering
\includegraphics{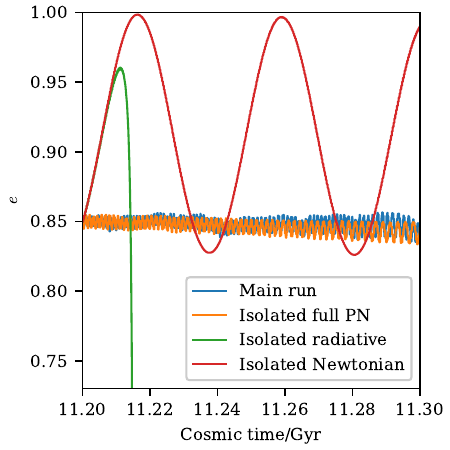}
\caption{
Evolution of the eccentricity $e$ for the inner AC binary during a part of
the phase where the system is in a hierarchical triplet configuration.
The evolution is shown for the full cosmological run (``main run'') as well as isolated
integrations of the triplet starting from the state at cosmic time $t={11.2\,\mathrm{Gyr}}$
using either the full PN equations of motion, including just the leading \pn{2.5}
radiative reaction term or using only Newtonian gravity.
}
\label{fig:kozai_eccentricity}
\end{figure}

Galaxy B merges with galaxy A at a redshift of ${z\approx 0.48}$.
In the \ketju{} simulation the SMBH of galaxy B sinks to the center of the merger remnant and forms a
binary with SMBH-A (AB-binary) with a semimajor axis of $a_\mathrm{AB} \approx {100\,\mathrm{pc}}$.
Over the following ${\sim 250 \, \mathrm{Myr}}$ stellar scattering hardens the binary
to a semimajor axis of $a_\mathrm{AB} \approx {10 \,\mathrm{pc}}$ 
(\autoref{fig:orbital_evolution}).
During this time period, galaxy C merges with the system as well, which results in 
a three-body interaction between the three SMBHs as SMBH-C sinks to the 
center of the system.
Initially this interaction causes rapid changes in the eccentricity of the 
AB-binary,
and finally SMBH-B is ejected from the center with SMBH-C
taking its place in a new binary with SMBH-A.

After a few hundred megayears, SMBH-B interacts again with the AC-binary, which
can be seen \edit1{from the small SMBH separations and the} dip in the AC eccentricity in 
\autoref{fig:orbital_evolution}.
This interaction ejects SMBH-B to an even wider orbit in the galaxy
(\autoref{fig:ejected_bh_orbit}),
from which it takes around a gigayear for it to sink back to the center of the galaxy.
Meanwhile, the AC-binary hardens due to stellar scattering, and finally merges
due to GW emission $\sim 3\,\mathrm{Gyr}$ after the galaxies merged.
The remaining AB-binary undergoes a similar evolution, but does not have time to
merge before $z=0$.

The eccentricity of AC shows small oscillations after B enters into a sub $\sim 100\,\mathrm{pc}$
hierarchical configuration.
\autoref{fig:kozai_eccentricity} shows these oscillations
during a period of time when
the inner binary has a semimajor axis of $a_\mathrm{AC}\approx{0.4\,\mathrm{pc}}$,
while SMBH-B is on an orbit of $a_\mathrm{AC-B}\approx{20\,\mathrm{pc}}$ with
eccentricity $e_\mathrm{AC-B}\approx 0.79$
at an inclination of $i_\mathrm{AC-B} \approx 90.8^{\circ}$.
The oscillations are what remains of Lidov--Kozai oscillations \citep{1962P&SS....9..719L} after being suppressed by the relativistic precession of the
inner orbit, due to the binary precession period
($\sim 6\times 10^5\,\mathrm{yr}$) being much shorter than the Lidov--Kozai oscillation
period ($\sim 4\times 10^7\,\mathrm{yr}$)
\citep{1997Natur.386..254H,2002ApJ...578..775B, 2016MNRAS.461.4419B}.
A comparison to an isolated integration of the system using only Newtonian gravity
shows that the system would indeed undergo large eccentricity oscillations without
the inclusion of relativistic precession from the \pn{1} level corrections.
With the addition of only the gravitational radiation reaction terms
the inner binary would merge rapidly due to these oscillations,
which serves to illustrate that the added complexity of the other
PN correction terms is necessary for correctly handling BH triplets or even more
complex systems.

Our \ketju{} PN correction implementation includes 
only PN terms relevant for binaries, ignoring three-body cross terms appearing
at \pn{1} level \citep[e.g.][]{1985PhRvD..31.1815T}.
It has been argued that these terms can in some cases lead to significant effects over long
enough time periods \citep{2014PhRvD..89d4043W,2020PhRvD.102f4033L}.
However, in this specific case the ignored terms do not appear to lead to significant
changes in the behavior of the system.
This is demonstrated by an isolated integration of the SMBH triplet using a version of the integrator
including also the \pn{1} level three-body terms, shown in  \autoref{fig:kozai_eccentricity}.
The results are visually almost indistinguishable from the main run,  
and utilizing additional integrations without the three-body terms we have confirmed
that the small differences between the runs are due to stellar
interactions.
\edit1{
However, in some other triplet configurations the cross terms may result in more significant effects,
and thus including them in future simulations seems prudent.
}

\subsection{SMBH Binary Hardening Rate}
To confirm that the SMBH binary hardening process is modeled correctly in a cosmological simulation, when 
including also stellar softening in the regularized regions, we fit the binary hardening rate using the
\cite{1996NewA....1...35Q} model
\begin{equation}
    \dv{a^{-1}}{t} = (a K)^{-1} \dv{e}{t} = H \frac{G\rho}{\sigma},
\end{equation}
where the stellar density $\rho$ and velocity dispersion $\sigma$ are computed
within the influence radius $R_\mathrm{inf} \approx 500\,\mathrm{pc}$
of the binary and $K$ and $H$ are constants. 
Performing the fit when the binary semimajor is around $a\sim 2\,\mathrm{pc}$,
we get for the AC binary at cosmic time $t\approx 10.4\,\mathrm{Gyr}$ the values
$H\approx 12$, $K\approx 0.1$, and for the AB binary at $t\approx 12.4\,\mathrm{Gyr}$
the slightly lower values $H\approx 5.2$, $K\approx 0.02$.
These results are comparable to the values obtained for our isolated elliptical galaxy
merger simulations \citep{2019ApJ...887...35M}. 
Based on these fits and using also analytical expressions for the effects of
GW emission \citep{1964PhRv..136.1224P}, we find that the AB binary would merge within
${\sim 400 \,\mathrm{Myr}}$ after the end of the simulation.

\subsection{Effects on the Stellar Density}

\begin{figure}
\centering
\includegraphics{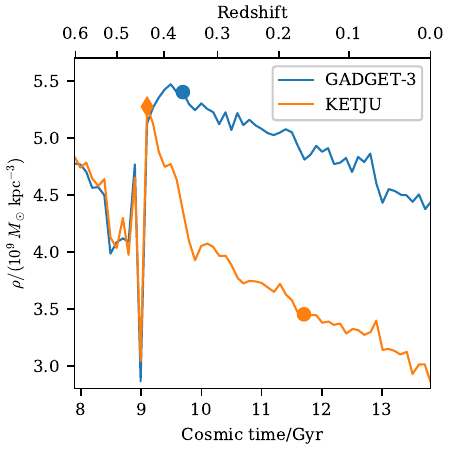}
\caption{
Evolution of the mean stellar density $\rho$ within a $r=500\,\mathrm{pc}$ sphere at the center of galaxy A.
The size of the sphere is approximately the same as the sphere of influence of the SMBH.
The circles mark  SMBH mergers, and the diamond indicates when a bound
binary was first formed in the \ketju{} run.
}
\label{fig:rho_evolution}
\end{figure}

With \ketju{} it is also possible to capture the effects of SMBH binaries on the
stellar distribution of the galaxies.
The evolution of the central stellar density around SMBH-A
is shown in \autoref{fig:rho_evolution}.
The \gadget{} run shows only a very gradual decrease after the galaxy mergers have
occurred and the SMBHs have merged at a separation of around one softening length.
In contrast, the \ketju{} run shows a rapid ejection of stars after the
formation of the bound SMBH system, tapering off to a more gradual decrease similar
to the \gadget{} run after ${\sim 1\,\mathrm{Gyr}}$.
The final stellar density of the \ketju{} run is lower by ${\sim 30\%}$, although 
the effect of the SMBH binaries on the stellar density in the \ketju{} run is not
quite as prominent as in some of our earlier isolated merger studies
\citep{2018ApJ...864..113R}, due to the lower masses of the SMBHs in the present study.

\section{Conclusions}

We have shown here how \ketju{} can be applied to cosmological zoom simulations
to capture the dynamics of massive SMBHs including also the complex behavior
of SMBH triplets.
 Our simulations are also able to resolve the effect of SMBH binaries on the distribution of stars in the central 
regions of the host galaxy, which can then affect the hardening rate and merger timescale 
of subsequently formed SMBH binaries in the same galaxy.
Modeling multiple SMBH systems 
and the detailed SMBH binary stellar interactions using only \edit1{simple orbit-averaged}
semi-analytic models is currently not feasible. The methods applied here can also be extended to study larger systems hosting tens of massive SMBHs,
with the main challenges including the relatively high required stellar mass resolution and the high computational 
cost of the regularized integration.  

The triple galaxy merger and the ensuing SMBH interactions presented here
demonstrate some key dynamical processes that complicate the SMBH merger
process in such systems compared to simple binary systems.
First, strong three-body interactions resulted in the ejection of one SMBH
to a wide orbit where it spent several gigayears.
In a system, which is sufficiently gas-rich, such an ejected SMBH could potentially be observable
as an offset AGN long after the initial galaxy mergers.
At a later stage the SMBH triplet entered a hierarchical configuration,
but due to relativistic precession there were no significant effects on orbit or
the merger timescale of the inner binary.
Had the inner binary been on a wider orbit with slower precession, the eccentricity
oscillations caused by the Lidov--Kozai mechanism could have significantly sped up the merger,
as was the case in the comparison integration with the precession effects disabled.
Detailed modeling of SMBHs in their global environment is therefore a
crucial tool for understanding the evolution and final fate of systems hosting multiple SMBHs.

\begin{acknowledgments} 
\modulolinenumbers[100] 

M.M., P.H.J. and S.L. acknowledge the support 
by the European Research Council via ERC Consolidator Grant KETJU (no. 818930). 

T.N. acknowledges support from the Deutsche Forschungsgemeinschaft 
(DFG, German Research Foundation) under Germany's Excellence Strategy - EXC-2094 - 390783311 from the DFG Cluster of Excellence ``ORIGINS''.

The numerical simulations used computational resources provided by 
the CSC -- IT Center for Science, Finland.
\end{acknowledgments}

\software{
\ketju{} \citep{2017ApJ...840...53R,2020MNRAS.492.4131R},
\gadget{} \citep{2005MNRAS.364.1105S},
NumPy \citep{numpy},
SciPy \citep{scipy},
Matplotlib \citep{matplotlib},
pygad \citep{2020MNRAS.496..152R},
MUSIC \citep{2011MNRAS.415.2101H,2013ascl.soft11011H}
}

\bibliographystyle{aasjournal}
\bibliography{refs}

\begin{thebibliography}{}
\expandafter\ifx\csname natexlab\endcsname\relax\def\natexlab#1{#1}\fi
\providecommand{\url}[1]{\href{#1}{#1}}
\providecommand{\dodoi}[1]{doi:~\href{http://doi.org/#1}{\nolinkurl{#1}}}
\providecommand{\doeprint}[1]{\href{http://ascl.net/#1}{\nolinkurl{http://ascl.net/#1}}}
\providecommand{\doarXiv}[1]{\href{https://arxiv.org/abs/#1}{\nolinkurl{https://arxiv.org/abs/#1}}}

\bibitem[{{Aumer} {et~al.}(2013){Aumer}, {White}, {Naab}, \&
  {Scannapieco}}]{2013MNRAS.434.3142A}
{Aumer}, M., {White}, S. D.~M., {Naab}, T., \& {Scannapieco}, C. 2013, \mnras,
  434, 3142, \dodoi{10.1093/mnras/stt1230}

\bibitem[{{Begelman} {et~al.}(1980){Begelman}, {Blandford}, \&
  {Rees}}]{begelman1980}
{Begelman}, M.~C., {Blandford}, R.~D., \& {Rees}, M.~J. 1980, \nat, 287, 307,
  \dodoi{10.1038/287307a0}

\bibitem[{{Blaes} {et~al.}(2002){Blaes}, {Lee}, \&
  {Socrates}}]{2002ApJ...578..775B}
{Blaes}, O., {Lee}, M.~H., \& {Socrates}, A. 2002, \apj, 578, 775,
  \dodoi{10.1086/342655}

\bibitem[{{Bonetti} {et~al.}(2016){Bonetti}, {Haardt}, {Sesana}, \&
  {Barausse}}]{2016MNRAS.461.4419B}
{Bonetti}, M., {Haardt}, F., {Sesana}, A., \& {Barausse}, E. 2016, \mnras, 461,
  4419, \dodoi{10.1093/mnras/stw1590}

\bibitem[{{Bonetti} {et~al.}(2019){Bonetti}, {Sesana}, {Haardt}, {Barausse}, \&
  {Colpi}}]{2019MNRAS.486.4044B}
{Bonetti}, M., {Sesana}, A., {Haardt}, F., {Barausse}, E., \& {Colpi}, M. 2019,
  \mnras, 486, 4044, \dodoi{10.1093/mnras/stz903}

\bibitem[{{Chandrasekhar}(1943)}]{1943ApJ....97..255C}
{Chandrasekhar}, S. 1943, \apj, 97, 255, \dodoi{10.1086/144517}

\bibitem[{{Deane} {et~al.}(2014){Deane}, {Paragi}, {Jarvis}, {Coriat},
  {Bernardi}, {Fender}, {Frey}, {Heywood}, {Kl{\"o}ckner}, {Grainge}, \&
  {Rumsey}}]{2014Natur.511...57D}
{Deane}, R.~P., {Paragi}, Z., {Jarvis}, M.~J., {et~al.} 2014, \nat, 511, 57,
  \dodoi{10.1038/nature13454}

\bibitem[{{Eisenreich} {et~al.}(2017){Eisenreich}, {Naab}, {Choi}, {Ostriker},
  \& {Emsellem}}]{2017MNRAS.468..751E}
{Eisenreich}, M., {Naab}, T., {Choi}, E., {Ostriker}, J.~P., \& {Emsellem}, E.
  2017, \mnras, 468, 751, \dodoi{10.1093/mnras/stx473}

\bibitem[{{Hahn} \& {Abel}(2011)}]{2011MNRAS.415.2101H}
{Hahn}, O., \& {Abel}, T. 2011, \mnras, 415, 2101,
  \dodoi{10.1111/j.1365-2966.2011.18820.x}

\bibitem[{{Hahn} \& {Abel}(2013)}]{2013ascl.soft11011H}
{Hahn}, O., \& {Abel}, T. 2013, {MUSIC: MUlti-Scale Initial Conditions}.
\newblock \doeprint{1311.011}

\bibitem[{{Harris} {et~al.}(2020){Harris}, {Millman}, {van der Walt},
  {Gommers}, {Virtanen}, {Cournapeau}, {Wieser}, {Taylor}, {Berg}, {Smith},
  {Kern}, {Picus}, {Hoyer}, {van Kerkwijk}, {Brett}, {Haldane}, {del R{\'\i}o},
  {Wiebe}, {Peterson}, {G{\'e}rard-Marchant}, {Sheppard}, {Reddy}, {Weckesser},
  {Abbasi}, {Gohlke}, \& {Oliphant}}]{numpy}
{Harris}, C.~R., {Millman}, K.~J., {van der Walt}, S.~J., {et~al.} 2020, \nat,
  585, 357, \dodoi{10.1038/s41586-020-2649-2}

\bibitem[{{Hills} \& {Fullerton}(1980)}]{1980AJ.....85.1281H}
{Hills}, J.~G., \& {Fullerton}, L.~W. 1980, \aj, 85, 1281,
  \dodoi{10.1086/112798}

\bibitem[{{Hoffman} \& {Loeb}(2007)}]{2007MNRAS.377..957H}
{Hoffman}, L., \& {Loeb}, A. 2007, \mnras, 377, 957,
  \dodoi{10.1111/j.1365-2966.2007.11694.x}

\bibitem[{{Holman} {et~al.}(1997){Holman}, {Touma}, \&
  {Tremaine}}]{1997Natur.386..254H}
{Holman}, M., {Touma}, J., \& {Tremaine}, S. 1997, \nat, 386, 254,
  \dodoi{10.1038/386254a0}

\bibitem[{{Hu} {et~al.}(2014){Hu}, {Naab}, {Walch}, {Moster}, \&
  {Oser}}]{2014MNRAS.443.1173H}
{Hu}, C.-Y., {Naab}, T., {Walch}, S., {Moster}, B.~P., \& {Oser}, L. 2014,
  \mnras, 443, 1173, \dodoi{10.1093/mnras/stu1187}

\bibitem[{{Hunter}(2007)}]{matplotlib}
{Hunter}, J.~D. 2007, Computing in Science and Engineering, 9, 90,
  \dodoi{10.1109/MCSE.2007.55}

\bibitem[{{Johansson} {et~al.}(2009{\natexlab{a}}){Johansson}, {Burkert}, \&
  {Naab}}]{2009ApJ...707L.184J}
{Johansson}, P.~H., {Burkert}, A., \& {Naab}, T. 2009{\natexlab{a}}, \apjl,
  707, L184, \dodoi{10.1088/0004-637X/707/2/L184}

\bibitem[{{Johansson} {et~al.}(2009{\natexlab{b}}){Johansson}, {Naab}, \&
  {Burkert}}]{2009ApJ...690..802J}
{Johansson}, P.~H., {Naab}, T., \& {Burkert}, A. 2009{\natexlab{b}}, \apj, 690,
  802, \dodoi{10.1088/0004-637X/690/1/802}

\bibitem[{{Kelley} {et~al.}(2017{\natexlab{a}}){Kelley}, {Blecha}, \&
  {Hernquist}}]{2017MNRAS.464.3131K}
{Kelley}, L.~Z., {Blecha}, L., \& {Hernquist}, L. 2017{\natexlab{a}}, \mnras,
  464, 3131, \dodoi{10.1093/mnras/stw2452}

\bibitem[{{Kelley} {et~al.}(2017{\natexlab{b}}){Kelley}, {Blecha}, {Hernquist},
  {Sesana}, \& {Taylor}}]{2017MNRAS.471.4508K}
{Kelley}, L.~Z., {Blecha}, L., {Hernquist}, L., {Sesana}, A., \& {Taylor},
  S.~R. 2017{\natexlab{b}}, \mnras, 471, 4508, \dodoi{10.1093/mnras/stx1638}

\bibitem[{{Khan} {et~al.}(2016){Khan}, {Fiacconi}, {Mayer}, {Berczik}, \&
  {Just}}]{2016ApJ...828...73K}
{Khan}, F.~M., {Fiacconi}, D., {Mayer}, L., {Berczik}, P., \& {Just}, A. 2016,
  \apj, 828, 73, \dodoi{10.3847/0004-637X/828/2/73}

\bibitem[{{Kollatschny} {et~al.}(2020){Kollatschny}, {Weilbacher}, {Ochmann},
  {Chelouche}, {Monreal-Ibero}, {Bacon}, \& {Contini}}]{2020A&A...633A..79K}
{Kollatschny}, W., {Weilbacher}, P.~M., {Ochmann}, M.~W., {et~al.} 2020, \aap,
  633, A79, \dodoi{10.1051/0004-6361/201936540}

\bibitem[{{Kormendy} \& {Ho}(2013)}]{kormendy2013}
{Kormendy}, J., \& {Ho}, L.~C. 2013, \araa, 51, 511,
  \dodoi{10.1146/annurev-astro-082708-101811}

\bibitem[{{Lidov}(1962)}]{1962P&SS....9..719L}
{Lidov}, M.~L. 1962, \planss, 9, 719, \dodoi{10.1016/0032-0633(62)90129-0}

\bibitem[{{Lim} \& {Rodriguez}(2020)}]{2020PhRvD.102f4033L}
{Lim}, H., \& {Rodriguez}, C.~L. 2020, \prd, 102, 064033,
  \dodoi{10.1103/PhysRevD.102.064033}

\bibitem[{{Liu} {et~al.}(2019){Liu}, {Hou}, {Li}, {Nyland}, {Guo}, {Kong},
  {Shen}, {Wrobel}, \& {Peng}}]{2019ApJ...887...90L}
{Liu}, X., {Hou}, M., {Li}, Z., {et~al.} 2019, \apj, 887, 90,
  \dodoi{10.3847/1538-4357/ab54c3}

\bibitem[{{Mannerkoski} {et~al.}(2019){Mannerkoski}, {Johansson}, {Pihajoki},
  {Rantala}, \& {Naab}}]{2019ApJ...887...35M}
{Mannerkoski}, M., {Johansson}, P.~H., {Pihajoki}, P., {Rantala}, A., \&
  {Naab}, T. 2019, \apj, 887, 35, \dodoi{10.3847/1538-4357/ab52f9}

\bibitem[{{Mikkola} \& {Merritt}(2008)}]{2008AJ....135.2398M}
{Mikkola}, S., \& {Merritt}, D. 2008, \aj, 135, 2398,
  \dodoi{10.1088/0004-6256/135/6/2398}

\bibitem[{{Mora} \& {Will}(2004)}]{2004PhRvD..69j4021M}
{Mora}, T., \& {Will}, C.~M. 2004, \prd, 69, 104021,
  \dodoi{10.1103/PhysRevD.69.104021}

\bibitem[{{Peters}(1964)}]{1964PhRv..136.1224P}
{Peters}, P.~C. 1964, Physical Review, 136, 1224,
  \dodoi{10.1103/PhysRev.136.B1224}

\bibitem[{{Peters} \& {Mathews}(1963)}]{1963PhRv..131..435P}
{Peters}, P.~C., \& {Mathews}, J. 1963, Physical Review, 131, 435,
  \dodoi{10.1103/PhysRev.131.435}

\bibitem[{{Pfeifle} {et~al.}(2019){Pfeifle}, {Satyapal}, {Manzano-King},
  {Cann}, {Sexton}, {Rothberg}, {Canalizo}, {Ricci}, {Blecha}, {Ellison},
  {Gliozzi}, {Secrest}, {Constantin}, \& {Harvey}}]{2019ApJ...883..167P}
{Pfeifle}, R.~W., {Satyapal}, S., {Manzano-King}, C., {et~al.} 2019, \apj, 883,
  167, \dodoi{10.3847/1538-4357/ab3a9b}

\bibitem[{{Pfister} {et~al.}(2019){Pfister}, {Volonteri}, {Dubois}, {Dotti}, \&
  {Colpi}}]{2019MNRAS.486..101P}
{Pfister}, H., {Volonteri}, M., {Dubois}, Y., {Dotti}, M., \& {Colpi}, M. 2019,
  \mnras, 486, 101, \dodoi{10.1093/mnras/stz822}

\bibitem[{{Planck Collaboration} {et~al.}(2020){Planck Collaboration},
  {Aghanim}, {Akrami}, {Ashdown}, {Aumont}, {Baccigalupi}, {Ballardini},
  {Banday}, {Barreiro}, {Bartolo}, {Basak}, {Battye}, {Benabed}, {Bernard},
  {Bersanelli}, {Bielewicz}, {Bock}, {Bond}, {Borrill}, {Bouchet}, {Boulanger},
  {Bucher}, {Burigana}, {Butler}, {Calabrese}, {Cardoso}, {Carron},
  {Challinor}, {Chiang}, {Chluba}, {Colombo}, {Combet}, {Contreras}, {Crill},
  {Cuttaia}, {de Bernardis}, {de Zotti}, {Delabrouille}, {Delouis}, {Di
  Valentino}, {Diego}, {Dor{\'e}}, {Douspis}, {Ducout}, {Dupac}, {Dusini},
  {Efstathiou}, {Elsner}, {En{\ss}lin}, {Eriksen}, {Fantaye}, {Farhang},
  {Fergusson}, {Fernandez-Cobos}, {Finelli}, {Forastieri}, {Frailis},
  {Fraisse}, {Franceschi}, {Frolov}, {Galeotta}, {Galli}, {Ganga},
  {G{\'e}nova-Santos}, {Gerbino}, {Ghosh}, {Gonz{\'a}lez-Nuevo}, {G{\'o}rski},
  {Gratton}, {Gruppuso}, {Gudmundsson}, {Hamann}, {Handley}, {Hansen},
  {Herranz}, {Hildebrandt}, {Hivon}, {Huang}, {Jaffe}, {Jones}, {Karakci},
  {Keih{\"a}nen}, {Keskitalo}, {Kiiveri}, {Kim}, {Kisner}, {Knox},
  {Krachmalnicoff}, {Kunz}, {Kurki-Suonio}, {Lagache}, {Lamarre}, {Lasenby},
  {Lattanzi}, {Lawrence}, {Le Jeune}, {Lemos}, {Lesgourgues}, {Levrier},
  {Lewis}, {Liguori}, {Lilje}, {Lilley}, {Lindholm}, {L{\'o}pez-Caniego},
  {Lubin}, {Ma}, {Mac{\'\i}as-P{\'e}rez}, {Maggio}, {Maino}, {Mandolesi},
  {Mangilli}, {Marcos-Caballero}, {Maris}, {Martin}, {Martinelli},
  {Mart{\'\i}nez-Gonz{\'a}lez}, {Matarrese}, {Mauri}, {McEwen}, {Meinhold},
  {Melchiorri}, {Mennella}, {Migliaccio}, {Millea}, {Mitra},
  {Miville-Desch{\^e}nes}, {Molinari}, {Montier}, {Morgante}, {Moss}, {Natoli},
  {N{\o}rgaard-Nielsen}, {Pagano}, {Paoletti}, {Partridge}, {Patanchon},
  {Peiris}, {Perrotta}, {Pettorino}, {Piacentini}, {Polastri}, {Polenta},
  {Puget}, {Rachen}, {Reinecke}, {Remazeilles}, {Renzi}, {Rocha}, {Rosset},
  {Roudier}, {Rubi{\~n}o-Mart{\'\i}n}, {Ruiz-Granados}, {Salvati}, {Sandri},
  {Savelainen}, {Scott}, {Shellard}, {Sirignano}, {Sirri}, {Spencer},
  {Sunyaev}, {Suur-Uski}, {Tauber}, {Tavagnacco}, {Tenti}, {Toffolatti},
  {Tomasi}, {Trombetti}, {Valenziano}, {Valiviita}, {Van Tent}, {Vibert},
  {Vielva}, {Villa}, {Vittorio}, {Wandelt}, {Wehus}, {White}, {White},
  {Zacchei}, \& {Zonca}}]{2020A&A...641A...6P}
{Planck Collaboration}, {Aghanim}, N., {Akrami}, Y., {et~al.} 2020, \aap, 641,
  A6, \dodoi{10.1051/0004-6361/201833910}

\bibitem[{{Quinlan}(1996)}]{1996NewA....1...35Q}
{Quinlan}, G.~D. 1996, \na, 1, 35, \dodoi{10.1016/S1384-1076(96)00003-6}

\bibitem[{{Rantala} {et~al.}(2018){Rantala}, {Johansson}, {Naab}, {Thomas}, \&
  {Frigo}}]{2018ApJ...864..113R}
{Rantala}, A., {Johansson}, P.~H., {Naab}, T., {Thomas}, J., \& {Frigo}, M.
  2018, \apj, 864, 113, \dodoi{10.3847/1538-4357/aada47}

\bibitem[{{Rantala} {et~al.}(2019){Rantala}, {Johansson}, {Naab}, {Thomas}, \&
  {Frigo}}]{2019ApJ...872L..17R}
{Rantala}, A., {Johansson}, P.~H., {Naab}, T., {Thomas}, J., \& {Frigo}, M.
  2019, \apjl, 872, L17, \dodoi{10.3847/2041-8213/ab04b1}

\bibitem[{{Rantala} {et~al.}(2017){Rantala}, {Pihajoki}, {Johansson}, {Naab},
  {Lah{\'e}n}, \& {Sawala}}]{2017ApJ...840...53R}
{Rantala}, A., {Pihajoki}, P., {Johansson}, P.~H., {et~al.} 2017, \apj, 840,
  53, \dodoi{10.3847/1538-4357/aa6d65}

\bibitem[{{Rantala} {et~al.}(2020){Rantala}, {Pihajoki}, {Mannerkoski},
  {Johansson}, \& {Naab}}]{2020MNRAS.492.4131R}
{Rantala}, A., {Pihajoki}, P., {Mannerkoski}, M., {Johansson}, P.~H., \&
  {Naab}, T. 2020, \mnras, 492, 4131, \dodoi{10.1093/mnras/staa084}

\bibitem[{{R{\"o}ttgers} {et~al.}(2020){R{\"o}ttgers}, {Naab}, {Cernetic},
  {Dav{\'e}}, {Kauffmann}, {Borthakur}, \& {Foidl}}]{2020MNRAS.496..152R}
{R{\"o}ttgers}, B., {Naab}, T., {Cernetic}, M., {et~al.} 2020, \mnras, 496,
  152, \dodoi{10.1093/mnras/staa1490}

\bibitem[{{Ryu} {et~al.}(2018){Ryu}, {Perna}, {Haiman}, {Ostriker}, \&
  {Stone}}]{2018MNRAS.473.3410R}
{Ryu}, T., {Perna}, R., {Haiman}, Z., {Ostriker}, J.~P., \& {Stone}, N.~C.
  2018, \mnras, 473, 3410, \dodoi{10.1093/mnras/stx2524}

\bibitem[{{Scannapieco} {et~al.}(2005){Scannapieco}, {Tissera}, {White}, \&
  {Springel}}]{2005MNRAS.364..552S}
{Scannapieco}, C., {Tissera}, P.~B., {White}, S.~D.~M., \& {Springel}, V. 2005,
  \mnras, 364, 552, \dodoi{10.1111/j.1365-2966.2005.09574.x}

\bibitem[{{Scannapieco} {et~al.}(2006){Scannapieco}, {Tissera}, {White}, \&
  {Springel}}]{2006MNRAS.371.1125S}
{Scannapieco}, C., {Tissera}, P.~B., {White}, S.~D.~M., \& {Springel}, V. 2006,
  \mnras, 371, 1125, \dodoi{10.1111/j.1365-2966.2006.10785.x}

\bibitem[{{Sijacki} {et~al.}(2007){Sijacki}, {Springel}, {Di Matteo}, \&
  {Hernquist}}]{2007MNRAS.380..877S}
{Sijacki}, D., {Springel}, V., {Di Matteo}, T., \& {Hernquist}, L. 2007,
  \mnras, 380, 877, \dodoi{10.1111/j.1365-2966.2007.12153.x}

\bibitem[{{Springel}(2005)}]{2005MNRAS.364.1105S}
{Springel}, V. 2005, \mnras, 364, 1105,
  \dodoi{10.1111/j.1365-2966.2005.09655.x}

\bibitem[{{Springel} {et~al.}(2005){Springel}, {Di Matteo}, \&
  {Hernquist}}]{2005MNRAS.361..776S}
{Springel}, V., {Di Matteo}, T., \& {Hernquist}, L. 2005, \mnras, 361, 776,
  \dodoi{10.1111/j.1365-2966.2005.09238.x}

\bibitem[{{Thorne} \& {Hartle}(1985)}]{1985PhRvD..31.1815T}
{Thorne}, K.~S., \& {Hartle}, J.~B. 1985, \prd, 31, 1815,
  \dodoi{10.1103/PhysRevD.31.1815}

\bibitem[{{Tremmel} {et~al.}(2015){Tremmel}, {Governato}, {Volonteri}, \&
  {Quinn}}]{2015MNRAS.451.1868T}
{Tremmel}, M., {Governato}, F., {Volonteri}, M., \& {Quinn}, T.~R. 2015,
  \mnras, 451, 1868, \dodoi{10.1093/mnras/stv1060}

\bibitem[{{Virtanen} {et~al.}(2020){Virtanen}, {Gommers}, {Oliphant},
  {Haberland}, {Reddy}, {Cournapeau}, {Burovski}, {Peterson}, {Weckesser},
  {Bright}, {van der Walt}, {Brett}, {Wilson}, {Millman}, {Mayorov}, {Nelson},
  {Jones}, {Kern}, {Larson}, {Carey}, {Polat}, {Feng}, {Moore}, {VanderPlas},
  {Laxalde}, {Perktold}, {Cimrman}, {Henriksen}, {Quintero}, {Harris},
  {Archibald}, {Ribeiro}, {Pedregosa}, {van Mulbregt}, \& {SciPy 1. 0
  Contributors}}]{scipy}
{Virtanen}, P., {Gommers}, R., {Oliphant}, T.~E., {et~al.} 2020, Nature
  Methods, 17, 261, \dodoi{10.1038/s41592-019-0686-2}

\bibitem[{{Will}(2014)}]{2014PhRvD..89d4043W}
{Will}, C.~M. 2014, \prd, 89, 044043, \dodoi{10.1103/PhysRevD.89.044043}

\end{thebibliography}

\end{document}